\documentclass[aps,pra,twocolumn,showpacs]{revtex4-1}
\usepackage{graphicx}
\usepackage{bm}
\usepackage{amsmath}
\usepackage{amssymb}
\usepackage{euscript}
\usepackage{verbatim}
\usepackage{fancyhdr}
\usepackage{wrapfig}
\usepackage{setspace}
\usepackage{xcolor}
\usepackage{amsfonts}
\usepackage{subfigure}

\newcommand{\Renyi}[0]{R\'{e}nyi}
\newcommand{\Tr}{\mathrm{Tr}}
\newcommand{\Tt}{{\mathrm{T}_2}}
\newcommand{\ket}[1]{\left| #1 \right\rangle}
\newcommand{\bra}[1]{\left\langle #1 \right|}
\newcommand{\tket}[1]{| #1 \rangle}
\newcommand{\tbra}[1]{\langle #1 |}

\newcommand{\mytitle}{Imbalance Entanglement: Symmetry Decomposition of Negativity}

\begin{document}

\title{\mytitle}

\author{Eyal Cornfeld}
\affiliation{Raymond and Beverly Sackler School of Physics and Astronomy, Tel-Aviv University, 6997801 Tel Aviv, Israel}

\author{Moshe Goldstein}
\affiliation{Raymond and Beverly Sackler School of Physics and Astronomy, Tel-Aviv University, 6997801 Tel Aviv, Israel}

\author{Eran Sela}
\affiliation{Raymond and Beverly Sackler School of Physics and Astronomy, Tel-Aviv University, 6997801 Tel Aviv, Israel}


\begin{abstract}
In the presence of symmetry, entanglement measures of quantum many-body states can be decomposed into contributions arising from distinct symmetry sectors. Here we investigate the decomposability of negativity, a measure of entanglement between two parts of a generally open system in a mixed state. While the entanglement entropy of a subsystem within a closed system can be resolved according to its total preserved charge, we find that negativity of two subsystems may be decomposed into contributions associated with their charge imbalance. We show that this charge-imbalance decomposition of the negativity may be measured by employing existing techniques based on creation and manipulation of many-body twin or triple states in cold atomic setups. Next, using a geometrical construction in terms of an Aharonov-Bohm-like flux inserted in a Riemann geometry, we compute this decomposed negativity in critical one-dimensional systems described by conformal field theory. We show that it shares the same distribution as the charge-imbalance between the two subsystems. We numerically confirm our field theory results via an exact calculations for non-interacting particles based on a double-gaussian representation of the partially transposed density matrix.
\end{abstract}

\maketitle

\section{Introduction and Results}\label{sec:int}

Negativity provides a measure of quantum entanglement between two subsystems $A_1$ and $A_2$ in a generally mixed state~\cite{Peres1996Separability,Rev2008Amico,LAFLORENCIE20161,PhysRevB.94.195121,PhysRevA.60.3496,lee2000partial,eisert1999comparison,PhysRevA.65.032314,PhysRevLett.95.090503,eisert2006entanglement}. This state can be achieved when the full system \({(A_1 \cup A_2) \cup B}\) is in a pure state, after tracing out \(B\), treated as the ``environment". In this case the usual von-Neumann entropy of either \(A_1\) or \(A_2\) 
is not a measure of quantum entanglement and instead other measures have to be defined such as the negativity. The latter involves the non-standard operation of a \emph{partial transposition} on the density matrix ${\rho_A \mapsto \rho_A^{\Tt}}$. A density matrix \(\rho_A\) is decomposable if \({\rho_A = \sum_{i} w^{\scriptscriptstyle(i)}_{\phantom{A_|}} \rho_{A_1}^{\scriptscriptstyle(i)}\otimes\rho_{A_2}^{\scriptscriptstyle(i)}}\), where \({\sum_{i} w^{\scriptscriptstyle(i)} =1}\) and \(\rho_{A_1},\rho_{A_2}\) are (positive) density matrices of the two subsystems; after performing a partial transposition on a decomposable density matrix with respect to the subsystem \(A_2\), all eigenvalues of \({\rho_{A_1}^{\scriptscriptstyle(i)}\otimes\rho_{A_2}^{\scriptscriptstyle(i)}}\) remain unchanged. One thus concludes that the presence of any negative eigenvalue in the spectrum \(\{\lambda\}\) of $\rho_A^{\Tt}$, referred to  as the negativity spectrum, must indicate entanglement between the two subsystems~\cite{Peres1996Separability}. One hence defines the entanglement negativity as
\begin{equation}
\mathcal{N}\equiv\frac{1}{2}\left(\Tr\big|\rho_A^{\Tt}\big|-1\right)=\sum_{\lambda<0}|\lambda|,
\end{equation}
such that non-vanishing negativity implies entanglement. Related entanglement measures are the \Renyi{} negativities,
\begin{equation}
R_n\equiv\Tr\{(\rho_A^{\Tt})^n\}=\sum_\lambda \lambda^n,\quad\; \mathcal{N}=\lim_{n\to 1/2}\tfrac{1}{2}\left(R_{2n}-1\right).
\end{equation}
Knowledge of \Renyi{} negativities may be used, via various techniques, to find either the entanglement negativity~\cite{gray2017measuring} or the entire negativity spectrum~\cite{Ruggiero2016Negativity}.

Over recent years there has been  growing interest in the negativity of many body systems. Key progress was achieved using field theory methods specifically focusing on critical systems~\cite{calabrese2012negativity,calabrese2013entanglement,calabrese2014finite,hoogeveen2015entanglement,coser2016towards,Ruggiero2016Negativity}, supplemented by numerical techniques~\cite{PhysRevB.90.064401,Ruggiero2016Negativity},  but also interesting aspects of negativity in topological gapped phases were discussed~\cite{PhysRevA.88.042318,PhysRevA.88.042319,wen2016topological}. Owing to the non-standard operation of partial transposition, obtaining the negativity spectrum is challenging even for free-fermion systems~\cite{eisler2015partial,coser2016towards}.

In this paper we study a general symmetry-decomposition of the negativity. Recently, it has been shown that entanglement entropy
admits a charge decomposition which can be both computed and measured~\cite{laflorencie2014spin,goldstein2017symmetry}. This is based on a block diagonal form of the density matrix in the presence of symmetries, and allows to identify contributions of entanglement entropy from individual charge sectors. It is natural to ask whether negativity  admits a similar symmetry-decomposition. This is nontrivial due to the involved operation of partial transposition on the density matrix. 

 
We find that instead of a decomposition according to the total charge, negativity admits a resolution by the charge-imbalance in the two subsystems. This holds whenever there is a conserved extensive quantity \(\hat{O}\) in the joint Hilbert space of the system \({A=A_1\cup A_2}\) and the environment \(B\), \textit{i.e.} \({\hat{O}_{(A_1\cup A_2)\cup B}=\hat{O}_{1}+\hat{O}_{2}+\hat{O}_{B}}\). We show that the negativity spectrum is then partitioned, \({\{\lambda_i\}=\bigcup_Q\{\lambda_{i_Q}\}}\), by the eigenvalues of an imbalance operator \({\hat{Q}\sim\hat{O}_1-\hat{O}_2}\); see Eq.~(\ref{maindec}) for the precise definition.
Examples for such extensive quantities may be the particle number \({\hat{N}_{(A_1\cup A_2)\cup B}=\hat{N}_{1}+\hat{N}_{2}+\hat{N}_{B}}\) or magnetization \({\hat{S}_{(A_1\cup A_2)\cup B}^z =\hat{S}_{1}^z+\hat{S}_{2}^z+\hat{S}_{B}^z}\). 

This finding
is particularly appealing in view of its experimental feasibility. Based on a proposal~\cite{alves2004multipartite,daley2012measuring} which had been experimentally implemented~\cite{Islam2015Measuring} to measure the \Renyi{} entanglement entropies of many-body states in cold atoms, a recent work~\cite{gray2017measuring} showed that the same experimental protocol can be simply generalized to measure the \Renyi{} negativity. Here, we demonstrate that a similar protocol naturally allows to measure separately the contributions to negativity from each symmetry sector.

The imbalance-decomposition of negativity is also compatible with the elegant field theory methods that were applied to compute the total negativity of critical 1D systems~\cite{calabrese2012negativity,calabrese2013entanglement,calabrese2014finite,hoogeveen2015entanglement,coser2016towards}. These computations are based on the replica trick approach, connecting the \Renyi{} negativities with the partition function of the theory on an $n$-sheeted Riemann surface connected via criss-cross escalators; see Fig.~\ref{fig:flux}. Interestingly, a proper insertion of an Aharonov-Bohm-like flux into this $n$-sheeted Riemann surface~\cite{Belin2013,Belin2015,Pastras2014,PhysRevB.93.195113,goldstein2017symmetry} can be used to obtain universal predictions for the imbalance-resolved negativity. We show that these contributions share the same distribution as that of the charge difference between subsystems $A_1$ and $A_2$. We numerically test our field theory calculations for free fermions~\cite{eisler2015partial} by mapping the partial transposed density matrix to a sum of two gaussian density matrices.

\begin{figure}[t]
\includegraphics[width=1\linewidth]{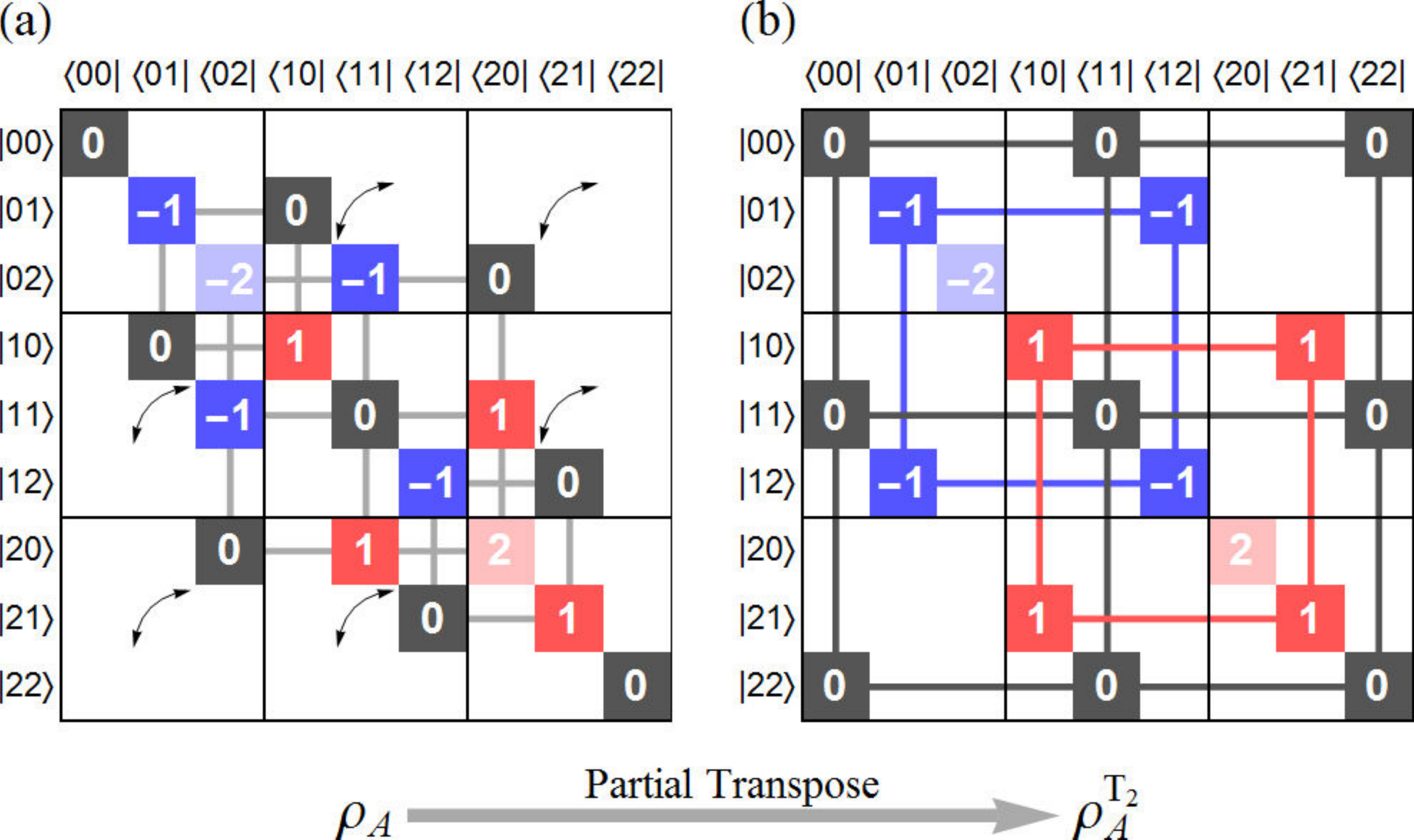}
\caption{(color online) Schematic block structure of density matrices in the basis of \(\tket{N_1N_2}\) with blocks labelled by \({Q=N_1-N_2^\Tt}\). (a) A density matrix \({\rho}_A\) has a block structure with respect to \({N_A=N_1+N_2}\) as shown by the thick lines. (b) Following a partial transposition, \({\rho}_A^\Tt\) has a block structure determined by \({Q=N_1-N_2^\Tt}\) as shown by the thick lines. The partial transposition is indicated by the arrows in (a), and relates the blocks according to Eq.~(\ref{blocks1}); for a specific example of these matrices see  Eqs.~(\ref{ex1}),~(\ref{ex1t2}).\label{fig:blocks}} 
\end{figure}

The plan of the paper is as follows. In Sec.~\ref{sec:gen}, we begin by motivating our work by a simple example. As illustrated in Fig.~\ref{fig:blocks}, while number conservation is reflected by a block diagonal structure of the density matrix, the operation of partial transposition mixes up these blocks; however, a new block structure is seen to emerge in terms of the imbalance operator $Q$. We proceed by a general definition of the imbalance operator and the associated decomposition of the negativity.
In Sec.~\ref{sec:exp}, we present an experimental protocol that enables  measurements of the resolved \Renyi{} negativities \([R_n]_Q\).
In Sec.~\ref{sec:cft} we focus on critical 1D systems and generalize conformal field theory (CFT) methods to derive a general result for the partition of the entanglement negativity.
We then test our field theory results by performing an exact numerical calculation for a free system. Finally, conclusions are provided in Sec.~\ref{sec:conclusion}.


\section{Imbalance Entanglement}\label{sec:gen}
In this section we provide a general definition of the symmetry resolution of entanglement negativity, referred to as ``imbalance entanglement". The key delicate issue to be addressed is the operation of \emph{partial transposition} of the density matrix, which is best demonstrated by a simple example.  

\subsection{Intuitive Example}\label{sec:ex}
In order to illustrate how symmetry is reflected in a block structure of the density matrix after partial transposition, it is  beneficial to begin with the simplest example. Consider a single particle located in one out of three boxes \(A_1, A_2, B\). It is described by a pure state \({\tket{\Psi} = \alpha \tket{100}+\beta \tket{010} +\gamma \tket{001}}\). The reduced density matrix for subsystem \({A=A_1 \cup A_2}\) is \({\rho_A = \Tr_B \tket{\Psi}\tbra{\Psi}} = {|\gamma|^2\tket{00}\tbra{00}+(\alpha \tket{10} + \beta \tket{01})(\alpha^\ast \tbra{10} + \beta^\ast \bra{01})}\), whose matrix representation is given by
\begin{equation}\label{ex1}
\rho_A=\left(\begin{array}{cc|cc}
	|\gamma|^2 & 0 &0 & 0\\
	0 & |\beta|^2 & \alpha^\ast \beta & 0\\
\hline
	0 & \beta^\ast\alpha & |\alpha|^2 & 0\\
	0 & 0 & 0 & 0
\end{array}\right),
\end{equation}
in the basis of \({\{\tket{00},\tket{01},\tket{10},\tket{11}\}}\). This matrix has a block diagonal structure with respect to the total occupation \({N_A=N_1+N_2}\),
\begin{equation}
\rho_A\cong\Big(\,|\gamma|^2\,\Big)_{N_A=0}\oplus\left(\begin{matrix}
	|\beta|^2 & \alpha^\ast \beta\\
	\beta^\ast \alpha & |\alpha|^2
\end{matrix}\right)_{N_A=1}\oplus\Big(\,0\,\Big)_{N_A=2}.
\end{equation}
Let us turn our attention to the partially transposed density matrix, \(\rho_A^{\Tt}\). It is obtained by transposing only the states of subsystem $A_2$ \textit{i.e.} \({\tket{N^{\phantom{.}}_1N^{\phantom{.}}_2}\tbra{N'_1N'_2}\mapsto\tket{N^{\phantom{.}}_1N'_2}\tbra{N'_1N^{\phantom{.}}_2}}\). This is equivalent to transposing the submatrices of \(\rho_A\),
\begin{equation}\label{ex1t2}
\rho_A^{\Tt}=\left(\begin{array}{cc|cc}
	|\gamma|^2 & 0 &0 & \alpha^\ast \beta\\
	0 & |\beta|^2 & 0 & 0\\
\hline
	0 & 0 & |\alpha|^2 & 0\\
	\beta^\ast\alpha & 0 & 0 & 0
\end{array}\right).
\end{equation}
The negativity spectrum for \(\rho_A^{\Tt}\) is easily found to be \({\{|\alpha|^2 ,  {\frac{1}{2}|\gamma|^2\pm\sqrt{\frac{1}{4}|\gamma|^4+|\alpha\beta|^2}} , |\beta|^2\}}\), and contains only one negative eigenvalue \({\mathcal{N}=\Big|\frac{1}{2}|\gamma|^2-\sqrt{\frac{1}{4}|\gamma|^4+|\alpha\beta|^2}\Big|}\).
Importantly, one may notice that \(\rho_A^{\Tt}\) has a block matrix structure. 
We label the blocks according to the occupation imbalance \({Q=N_1-N_2}\) of their diagonal elements,
\begin{equation}
\rho_A^{\Tt}\cong\Big(\,|\alpha|^2\,\Big)_{Q=1}\oplus\left(\begin{matrix}
	|\gamma|^2 & \alpha^\ast \beta\\
	\beta^\ast \alpha & 0
\end{matrix}\right)_{Q=0}\oplus\Big(\,|\beta|^2\,\Big)_{Q=-1}.
\end{equation}
Here, \({N_1-N_2=1}\) corresponds to \(\{\tket{10}\}\), \({N_1-N_2=0}\) corresponds to \(\{\tket{00}, \tket{11}\}\), and \({N_1-N_2=-1}\) corresponds to \(\{\tket{01}\}\). 
This decomposition partitions the negativity spectrum \(\{|\alpha|^2\}\cup{\{ \frac{1}{2}|\gamma|^2\pm\sqrt{\frac{1}{4}|\gamma|^4+|\alpha|^2|\beta|^2}\}}\cup\{|\beta|^2\}\).
As we henceforth show, this partitioning of the negativity spectrum goes beyond this example and is applicable to the general case.

\subsection{General Definition}
One may define an imbalance partition with respect to any extensive operator \({\hat{O}_{(A_1\cup A_2)\cup B}=\hat{O}_{1}+\hat{O}_{2}+\hat{O}_{B}}\). For simplicity we focus on the case of a conserved total particle number \({\hat{N}_{(A_1\cup A_2)\cup B}=\hat{N}_{1}+\hat{N}_{2}+\hat{N}_{B}}\).
Let us explore the consequences of this conservation law. It is reflected in the relation  \({[\rho_A,\hat{N}_A]=0}\)  satisfied by the reduced density matrix ${\rho_A = \Tr_B \rho}$ (\emph{e.g.}, a thermal state \({\rho_A \propto e^{-\beta\hat{H}}}\)). Partially transposing this commutation relation yields
\begin{equation}
[\rho_A^{\Tt},\hat{N}_1^{\phantom{|}}-\hat{N}_2^{\Tt}]=0.
\end{equation}
This commutativity elicits a block matrix decomposition, 
\begin{equation}
\label{maindec}
\rho_A^{\Tt}=\bigoplus_{Q}[\rho_A^{\Tt}]_Q,\qquad \hat{Q}\equiv\hat{N}_1^{\phantom{|}}-\hat{N}_2^{\Tt},
\end{equation}	 
where \(Q\) are the eigenvalues of \(\hat{Q}\). 
It is easily verified that this resolution is basis independent \textit{i.e.} that the spectrum of \([\rho_A^{\Tt}]_Q\) is invariant to local basis transformations \({\hat{O}\mapsto(\hat{U}_{1}^\dag\hat{U}_{2}^\dag)\hat{O}(\hat{U}_{1}^{\phantom{\dagger}}\hat{U}_{2}^{\phantom{\dagger}})}\) for all transformations \(\hat{U}_\alpha\) acting only in regions \(A_\alpha\).
The negativity spectrum \(\{\lambda_i\}\) of \([\rho_A^{\Tt}]\) may thus be decomposed
\({\{\lambda_i\}=\bigcup_Q\{\lambda_{i_Q}\}}\)
into spectra \(\{\lambda_{i_Q}\}\) of \([\rho_A^{\Tt}]_Q\), such that the overall entanglement negativity is resolved into contributions from distinct imbalance sectors
\begin{equation}\label{resolve1}
\Tr\big|\rho_A^{\Tt}\big|=\sum_{Q}\Tr\big\{\hat{P}_Q\big|\rho_A^{\Tt}\big|\big\}=\sum_{Q}\Tr\big|[\rho_A^{\Tt}]_Q\big|,
\end{equation}
where \(\hat{P}_Q\) is the projector to the subspace of eigenvalue~\(Q\) of the operator $\hat{Q}$. Similarly the \Renyi{} negativity is decomposed as ${R_n=\sum_Q[R_n]_Q}$, where
\begin{equation}
[R_n]_Q\equiv\Tr\{\hat{P}_Q(\rho_A^{\Tt})^n\}=\Tr\{([\rho_A^{\Tt}]_Q)^n\}.
\end{equation}

Generalizing the example from the previous subsection,
we write the density matrix as
\begin{equation}
\label{notation}
\rho_A=\sum_{\{N\},\{i\}}\tket{N^{\phantom{.}}_1}^{i^{\phantom{.}}_1}\tket{N^{\phantom{.}}_2}^{i^{\phantom{.}}_2}[\rho_A]_{N^{\phantom{.}}_1,N^{\phantom{.}}_2;N'_1,N'_2}^{i^{\phantom{.}}_1,i^{\phantom{.}}_2;i'_1,i'_2}\tbra{N'_1}^{i'_1}\tbra{N'_2}^{i'_2},
\end{equation}
where \(\{\tket{N_\alpha}^1,\tket{N_\alpha}^2,\dots\}\) span the Hilbert space of \(A_\alpha\) with particle number \(N_\alpha\)~(${\alpha \in \{A_1,A_2\}}$). Charge conservation implies that $\rho_A$ commutes with $\hat{N}_1 +\hat{N}_2$. This leads to a block structure of $\rho_A$ with $N_1 + N_2 = N_1'+N_2'$. This block structure of the density matrix is illustrated in the Fig.~\ref{fig:blocks}(a). We now wish to see which of these blocks contribute to each imbalance sector.

The block structure of \(\rho_A\) with \({N^{\phantom{.}}_1+N^{\phantom{.}}_2=N'_1+N'_2}\) allows us to identify ${Q=N^{\phantom{.}}_1-N'_2=N^{\phantom{.}}_2-N'_1}$, and assign a specific value of $Q$ to each block, as marked inside the squares ($N^{\phantom{.}}_1,N^{\phantom{.}}_2;N'_1,N'_2$ blocks) in Fig.~\ref{fig:blocks}(a). For example, the coherence term, $\beta^*\alpha$, in Sec.~\ref{sec:ex} corresponds to ${|N^{\phantom{.}}_1 N^{\phantom{.}}_2 \rangle \langle N_1' N_2' | =|1 0 \rangle \langle 0 1 |}  $, and hence to ${Q=0}$. 

Now, consider the partially transposed density matrix. The blocks of the original density matrix that contribute to \([\rho_A^{\Tt}]_Q\) for a given $Q$ are
\begin{equation}\label{blocks1}
[\rho_A^\Tt]_{N_1,N_1-Q;N_2+Q,N_2}^{\{i\}}=[\rho_A]_{N_1,N_2;N_2+Q,N_1-Q}^{\{i\}}.
\end{equation}
As can be seen in Fig.~\ref{fig:blocks}(b), these blocks reorganize into a diagonal block structure labelled by $Q$ after partial transposition. Staring at diagonal blocks of $\rho_A$, \emph{i.e.},  ${(N^{\phantom{.}}_1 , N^{\phantom{.}}_2) = (N_1' , N_2')}$, we see that ${Q \leftrightarrow   N_1 -N_2}$ is just the charge imbalance between the two subsystems, motivating the term ``imbalance decomposition" [note, though, that the density matrix also contains non-diagonal blocks where $(N_1 , N_2) \ne (N_1' , N_2')$, and that the blocks that contribute to the $Q$ imbalance sectors are precisely those in Eq.~(\ref{blocks1})].
 
\section{Protocol for Experimental Detection}\label{sec:exp}
In the previous section we identified the imbalance operator $Q$ according to which the partially transposed density matrix admits a block-diagonal form, allowing to decompose the negativity spectrum.  In this section we  show that a measurement of the individual imbalance-contributions to the \Renyi{} negativities can be performed within an existing experimental setup. For this purpose we adopt protocols which have been recently implemented in an experiment measuring entanglement entropy~\cite{Islam2015Measuring}. Specifically, in order to measure the resolved \Renyi{} negativity, \([R_n]_Q\), we will build on a recent proposed protocol~\cite{gray2017measuring} based on Ref.~[\onlinecite{daley2012measuring}] designed specifically to measure the total \Renyi{} negativity \(R_n\).

We begin this section presenting the basic idea of the protocol for measuring entanglement entropy, and then progressively show how the entanglement entropy and the negativity can be measured and partitioned according to symmetry sectors. The impatient reader interested directly in the protocol may skip to Sec.~\ref{se:negpart}.

\begin{figure}[t]
\includegraphics[width=1\linewidth]{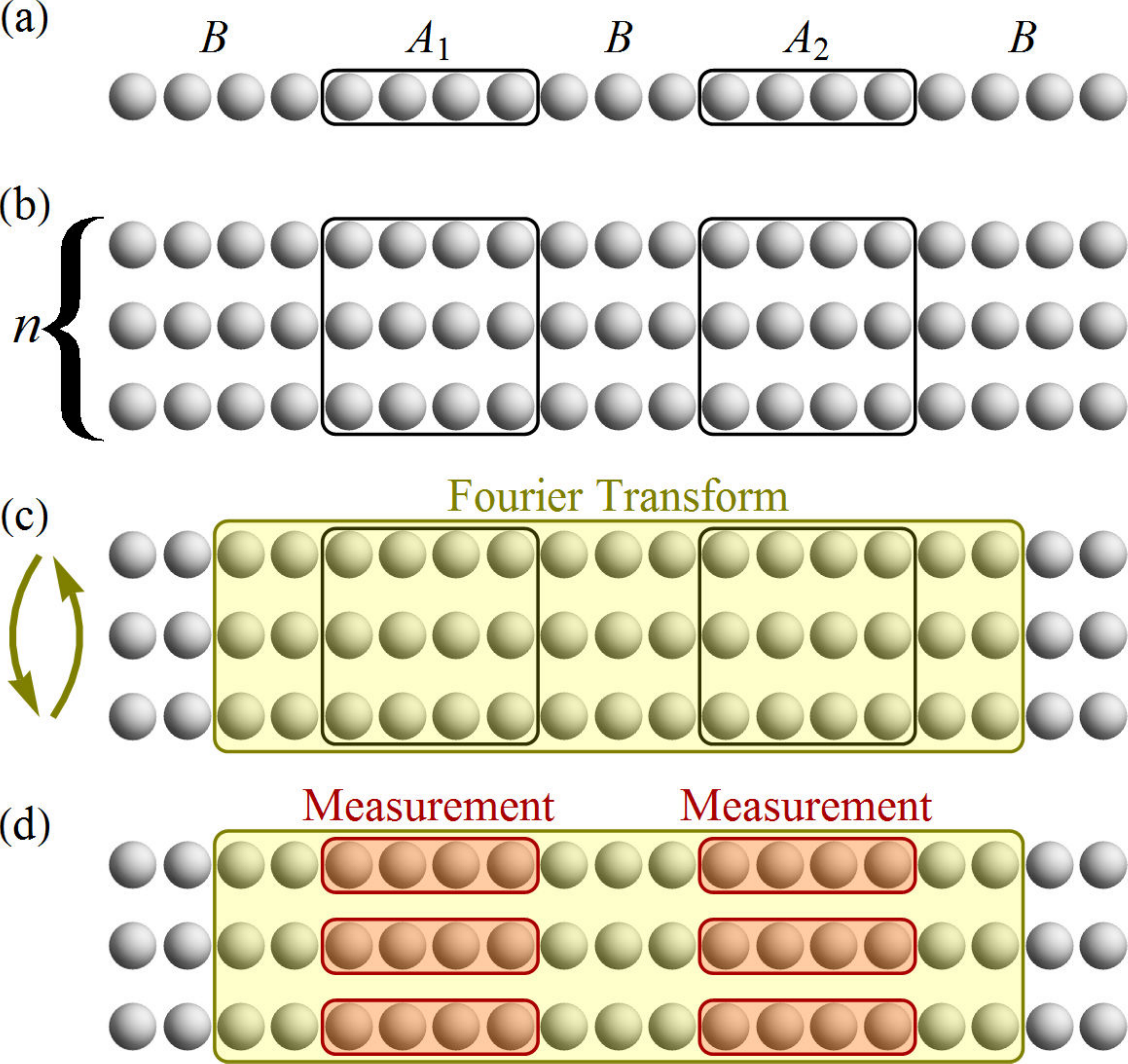}
\caption{(color online) Depiction of the experimental protocol to measure the imbalance-resolved \Renyi{}  negativity. (a) The system is tripartitioned into regions \(A_1,A_2,B\). (b) Prepare \(n\)-copies of the system. (c) Perform a copy-space Fourier transform on a region containing \(A_1,A_2\). (d) Measure \(N^{k}_{A_1},N^{k}_{A_2}\) in regions \(A_1,A_2\) (${k=1\dots n})$, and calculate \({Q=\frac{1}{n}\sum_k(N^{k}_{A_1}-N^{k}_{A_2})}\) and \({r_n=  \exp\{\sum_k\frac{2\pi i}{n}k(N^{k}_{A_1}-N^{k}_{A_2})\}}\) of Eq.~(\ref{QrQ}). The contribution of imbalance sector $Q$ to the \Renyi{} negativity is then given by the average over $\delta_{Q,q}\times r_n$. \label{fig:experiment}} 	 	
\end{figure} 

\subsection{Key Idea and the Swap Operator}
The starting point for the entanglement measurement protocols under consideration is a preparation of \(n\) copies of the many-body system. If, for instance, the original Hilbert space under consideration corresponds to that of a 1D chain as depicted in Fig.~\ref{fig:experiment}(a), then one extends the Hilbert space into a product of $n$ such Hilbert spaces describing $n$ identical chains as described in Fig.~\ref{fig:experiment}(b). In this space one wishes to prepare the $n$-copy state \(\rho_A^{\otimes n}\) of the original state $\rho_A$. This can be achieved~\cite{Islam2015Measuring} using optical lattices in cold atom systems, where one simulates the same Hamiltonian on the $n$ initially decoupled identical chains.

One then defines an operator in the extended Hilbert space, swapping between the quantum states of the copies in region $A$. It is simpler to restrict our attention to region $A$ from now on in this section. Explicitly, we denote a basis of states on the $n-$copy Hilbert space of region $A$ by \(\tket{ \psi^{1},\psi^{2},\dots,\psi^{n}}\), and the swap operator $\hat{S}$ is defined via
\begin{equation}
\hat{S}\tket{\psi^{1},\psi^{2},\dots,\psi^{n}}=\tket{\psi^{n},\psi^{1},\dots,\psi^{n-1}}.
\end{equation}
The key relation used in the protocol is that the \Renyi{} entropies, \(S_n\), satisfy
\begin{equation}
\label{eq:rel}
S_n\equiv\Tr\{\rho_A^n\}  =  \Tr\{\hat{S}\rho_A^{\otimes n}\},
\end{equation} 
\emph{i.e.}, the expectation value of the swap operator in the $n$-copy state equals the desired \Renyi{}  entropy~\cite{daley2012measuring}. As described in detail in the next subsection, the protocol proceeds by a proper manipulation of the $n$-copy system via a transformation between the copies, see Fig.~\ref{fig:experiment}(c), followed by site-resolved measurements, see Fig.~\ref{fig:experiment}(d). The combination of the latter two is designed precisely to implement a measurement of the expectation value of the swap operator in the $n$-copy state. 

\subsection{Measuring \Renyi{}  Entropy}
Before turning to negativity, it is convenient to introduce notation and show that the protocol just described indeed can measure the expectation value of the swap operator, and hence the \Renyi{}  entropy. Consider a general bosonic state in the occupation basis
\begin{equation}
|\vec{m}^{1},\vec{m}^{2}, \dots, \vec{m}^{n} \rangle = \prod_{k=1}^n\prod_{i\in A}\frac{([\hat{a}^{\dagger}]^{k}_{i})^{m^{k}_i}}{\sqrt{m^{k}_i!}}\tket{0}.
\end{equation}
Here, the index $k$ runs over the $n$ copies, and $i$ runs over all sites in region $A$.
We then perform a Fourier transform in the \(k=1\dots n\) copy-space 
\begin{equation}
\label{FT}
\hat{F}[\hat{a}^{\dagger}]^{k}_{i}\hat{F}^\dagger=\frac{1}{\sqrt{n}}\sum_{\ell=1}^n \omega^{k\ell}[\hat{a}^{\dagger}]^{\ell}_{i},
\end{equation}
where \({\omega=e^{\frac{2\pi i}{n}}}\), such that
\begin{equation}
\hat{F}|\vec{m}^{1}, \dots, \vec{m}^{n} \rangle= \prod_{k ,i}\frac{\Big( \frac{1}{\sqrt{n}}\sum_{\ell}\omega^{k\ell}[\hat{a}^{\dagger}]^{\ell}_{i} \Big)^{m^{k}_i}}{\sqrt{m^{k}_i!}}\tket{0}.
\end{equation}
Let us look at the operator
\begin{gather}
\label{Udef}
\hat{U}\equiv\omega^{\sum_{ k=1}^n k \hat{N}^{k}},
\end{gather}
where \({\hat{N}^{k}=\sum_{i\in A} [\hat{a}^{\dagger}]^{k}_{i}[\hat{a}]^{k}_{i}}\) is the total particle number operator in region $A$ in the \(k\)-th copy. It satisfies ${\hat{U}[\hat{a}^{\dagger}]^{k}_{i}=\omega^{k}[\hat{a}^{\dagger}]^{k}_{i}\hat{U}}$. Using this relation, when acting on the Fourier transformed state  this operator gives
\begin{align}
\label{steps}
&\hat{U}\hat{F}|\vec{m}^{1},\vec{m}^{2}, \dots, \vec{m}^{n} \rangle \nonumber\\ 
&=\prod_{k,i}\tfrac{1}{ \sqrt{m^{k}_i!}}\Big(\tfrac{1}{\sqrt{n}}\sum\nolimits_{\ell}\omega^{(k+1)\ell}[\hat{a}^{\dagger}]^{\ell}_{i}\Big)^{m^{k}_i} \tket{0}\nonumber\\
&=\prod_{k,i}\tfrac{1}{ \sqrt{m^{k}_i!}}\Big(\tfrac{1}{\sqrt{n}}\sum\nolimits_{\ell}\omega^{k\ell}[\hat{a}^{\dagger}]^{\ell}_{i}\Big)^{m^{k-1}_i} \tket{0}\nonumber\\
&=\hat{F}|\vec{m}^{n},\vec{m}^{1}, \dots ,\vec{m}^{n-1} \rangle=\hat{F}\hat{S}|\vec{m}^{1}, \dots, \vec{m}^{n} \rangle.
\end{align}
In exchanging the order of creation operators we have used the bosonic commutation relations. Since this equation holds for all states $|\vec{m}^{1},\vec{m}^{2}, \dots, \vec{m}^{n} \rangle$, one has
\begin{equation}
\label{keyrel}
\hat{F}^\dagger\hat{U}\hat{F}=\hat{S}.
\end{equation}
This operator identity implies that measurements of \({ \hat{U}=  \omega^{\sum_{k=1}^n k \hat{N}^{k}}}\) on the Fourier transformed system yields the expectation value of $\hat{S}$, hence, by Eq.~(\ref{eq:rel}), if the system is in an $n-$copy state, this gives the \Renyi{}  entropy ${S_n=\Tr\{\rho_A^n\}}$.

In other words, defining the set of commuting occupancies of region $A$ of the various copies after the Fourier transform by $\{ \tilde{N}\}$, where ${{{\hat{\tilde{N}}}{}^{k} = {\hat{F}^\dagger} {{\hat{N}}}^{k} {\hat{F}}}}$, the experiment simply performs a measurement in this new occupation basis and computes a function $f(\{ \tilde{N} \})$ of the outcomes $\{ \tilde{N} \}$ given by
\begin{equation}
f(\{ \tilde{N} \}) \equiv \omega^{\sum_{k=1}^n k \tilde{N}^{k}}.
\end{equation}
Operationally, the operational identity Eq.~(\ref{keyrel}) is equivalent to ${\hat{f}(\{ \tilde{N} \}) = \omega^{ \sum_{k}  { k\hat{\tilde{N}}}{}^{k}}=\hat{S}}$. Thus, we can describe the experimental protocol via
\begin{equation}
\label{keyequation}
\Tr \{ \rho_A^{\otimes n} \hat{f}(\{ \tilde{N} \} ) \}  = \Tr \{ \rho_A^{\otimes n} \hat{S} \} .
\end{equation}
The left hand side describes the measurement performed in the $\{ \tilde{N}^{k} \}$ basis, and the right hand side is the desired quantity.

\subsection{Measuring Charge-Resolved \Renyi{} Entropy}
\label{se:negmeasure}
Using this notation it becomes simple to demonstrate the protocol for measuring the charge-resolved \Renyi{}  entropy~\cite{goldstein2017symmetry}.

In the presence of a conserved number of particles $N_A$ the general density matrix can be written, using the same notation of Eq.~(\ref{notation}) as
\begin{equation}
\rho_A = \sum_N \sum_{i,i'} | N \rangle^i  [\rho_A]_N^{i i'} \langle N |^{i'}.
\end{equation}
Here, $| N \rangle^i$ and $| N \rangle^{i'}$ are different states having the same total number of particles $N$. Therefore, the $n$-copy density matrix can be written, using a super-index ${\tket{\mathbf{N}} \equiv \tket{N^{1},N^{2},\dots, N^{n}}^{i^{1},i^{2},\dots,i^{n}}}$, as
\begin{equation}
\label{rhoNN}
\rho_A^{\otimes n} = \sum_{\mathbf{N},\mathbf{N}'}  [\rho_A]_{\mathbf{N},\mathbf{N}'} | \mathbf{N} \rangle \langle \mathbf{N}' |,
\end{equation}
where ${N^{k} = N'^{k}}$ (${k=1\dots n}$). We are now interested in measuring the contribution of the charge ${N_A=N}$ block of the density matrix, ${[\rho_A]_N =  \sum_{i,i'} | N \rangle^i  [\rho_A]_N^{i i'} \langle N |^{i'}}$, to the \Renyi{}  entropy~\cite{goldstein2017symmetry}. This quantity ${[S_n]_N = \Tr\{([\rho_A]_N)^n\}}$, equals the expectation value of the swap operator in the \(n\)-copy state obtained by $[\rho_A]_N^{\otimes n}$. This may be measured after the Fourier transformation by calculating the function
\begin{equation}
f_{N_A}(\{ \tilde{N} \} ) \equiv  \delta_{N_A,\frac{1}{n}\sum_{k=1}^n \tilde{N}^{k}} \times \omega^{\sum_{k=1}^n k \tilde{N}^{k}}.
\end{equation}
To prove this statement we start with
\begin{align}
\label{step1}
\Tr \{ \rho_A^{\otimes n} \hat{f}_{N_A}( \{ \tilde{N} \} ) \}  =\sum_{\mathbf{N},\mathbf{N}'} [\rho_A]_{\mathbf{N},\mathbf{N}'}  \langle \mathbf{N}' |\hat{f}_{N_A}( \{ \tilde{N} \} )| \mathbf{N} \rangle.
\end{align}
Since the sum of the particle number is invariant under the Fourier transformation, ${\sum_{k=1}^n \tilde{N}^{k} = \sum_{k=1}^n N^{k}}$, Eq.~(\ref{step1}) equals
\begin{multline}
\label{step2}
\sum_{\mathbf{N},\mathbf{N}'}  (\delta_{N_A,\frac{1}{n}\sum_{k} N^{k}}) [\rho_A]_{\mathbf{N},\mathbf{N}'}  \langle \mathbf{N}' |\omega^{ \sum_{k} k \tilde{N}^{k}}| \mathbf{N} \rangle   \\
= \sum_{\mathbf{N},\mathbf{N}'} (\delta_{N_A,\frac{1}{n}\sum_{k} N^{k}}) [\rho_A]_{\mathbf{N},\mathbf{N}'}  \langle \mathbf{N}' |\hat{S}| \mathbf{N} \rangle,
\end{multline}
where the last equality uses Eq.~(\ref{keyrel}).
Crucially, the matrix element of the swap operator vanishes except if $N^{1}= N^{2} = \cdots =N^{n}$, namely it is proportional to $\prod_{k,k'} \delta_{N^{k},N^{k'}}$.  Thus, we conclude that 
\begin{align}
\label{step3}
\Tr \{ \rho_A^{\otimes n} \hat{f}_{N_A}( \{ \tilde{N} \} ) \}   &= \sum_{\mathbf{N},\mathbf{N}'} (\prod\limits_k \delta_{N_A,N^{k}}) [\rho_A]_{\mathbf{N},\mathbf{N}'}  \langle \mathbf{N}' |\hat{S}| \mathbf{N} \rangle \nonumber \\
&=\Tr \{ [\rho_A]_{N_A}^{\otimes n}\hat{S} \} =[S_n]_{N_A}.
\end{align}
\subsection{Measuring Negativity}
Switching to the \Renyi{} negativity, one defines a twisted swap operator \(\hat{T}=\hat{S}^{\phantom{1}}_{A_1}\hat{S}^{-1}_{A_2}\) on the \(n\)-copy Hilbert space of region \(A\), such that
\begin{equation}
\hat{T}\ket{\begin{aligned}&\vec{m}_{A_1}^{1},\dots,\vec{m}_{A_1}^{n}\\ &\vec{m}_{A_2}^{1},\dots,\vec{m}_{A_2}^{n}\end{aligned}}=\ket{\begin{aligned}&\vec{m}_{A_1}^{n}, \vec{m}_{A_1}^{1},\dots,\vec{m}_{A_1}^{n-1}\\ &\vec{m}_{A_2}^{2},\dots,\vec{m}_{A_2}^{n},\vec{m}_{A_2}^{1\phantom{-1}}\end{aligned}}.
\end{equation}
Here, 
this basis specifies the occupations $\vec{m}_{A_1}^{1} \dots \vec{m}_{A_1}^{n}$ on the $n$ copies of subsystem $A_1$ as well as the occupations  $\vec{m}_{A_2}^{1} \dots \vec{m}_{A_2}^{n}$ on the $n$ copies of subsystem $A_2$. In analogy with Eq.~(\ref{eq:rel}), the \Renyi{} negativity is given by the expectation value of this swap operator in the $n$-copy state~\cite{gray2017measuring},
\begin{equation}
R_n = \Tr\{(\rho_A^{\Tt})^n\}=\Tr\{\hat{T}\rho_A^{\otimes n}\}.
\end{equation}

To proceed with the measurement of the expectation value of this swap operator in the $n$-copy state, we use the same Fourier transform on all sites as in Eq.~(\ref{FT}), but generalize the definition of the $\hat{U}$ operator in Eq.~(\ref{Udef}) to
\begin{equation}
\hat{U} \equiv \omega^{\sum_{k=1}^n k (\hat{N}^{k}_{A_1} - \hat{N}^{k}_{A_2})}.
\end{equation}
Here, \(N^{k}_{A_\alpha}\) is the total particle number operator in subsystem \(A_\alpha\) of the \(k\)-th copy, \emph{i.e.}, ${N^{k}_{A_1} = \sum_{i \in A_1} m^{k}_i}$ and ${N^{k}_{A_2} = \sum_{i \in A_2} m^{k}_i}$. Following the same steps as in Eq.~(\ref{steps}), we readily obtain the relation
\begin{equation}
\label{EqT}
\hat{T}=\hat{F}^\dagger \omega^{ \sum_{k=1}^n k(\hat{N}^{k}_{A_1}-\hat{N}^{k}_{A_2})}\hat{F}.
\end{equation}
Thus, the physical protocol firstly consists of a unitary Hamiltonian evolution depicted in Fig.~\ref{fig:experiment}(c) which implements~\cite{daley2012measuring} the Fourier transformation $\hat{F}$. Then, as in Fig.~\ref{fig:experiment}(d), we perform a measurement of the  total occupancies $N^{k}_{A_1}$ and $N^{k}_{A_2}$ in each region $A_1$ and $A_2$ and in each copy, from which
\begin{equation}
f^{\mathrm{neg}}(\{N\})\equiv\omega^{\sum_{k=1}^n k(N^{k}_{A_1}-N^{k}_{A_2})}
\end{equation}
can be computed, and averaged over many realizations to obtain the total \Renyi{} negativity.

\subsection{Measuring Imbalance-Resolved Negativity}
\label{se:negpart} We are now ready to provide a protocol to measure the symmetry-resolved negativity. We begin with a step by step description of the protocol, followed by an outline of the proof which relies on the previous subsections.

Our protocol to measure the \Renyi{} negativity \([R_n]_Q\) for bosons consists of the following steps (see Fig.~\ref{fig:experiment}):\\
(i) Prepare \(n\) copies of the desired system.\\
(ii) Decouple the sites within each copy and perform a Fourier transform on every site between the copies. This is achieved by a unitary Hamiltonian evolution which implements Eq.~(\ref{FT}). \\
(iii) Perform a measurement of the total particle number \(N^{k}_{A_1}\) in subsystem \(A_1\) and of \(N^{k}_{A_1}\) in \(A_2\) for each copy \({k=1\dots n}\).\\
(iv) Calculate
\begin{equation}\label{QrQ}
q\equiv\frac{1}{n}\sum_{k=1}^n\left(N^{k}_{A_1}-N^{k}_{A_2}\right),\quad r_n\equiv e^{\frac{2\pi i}{n} \sum_{k=1}^{n}k(N^{k}_{A_1}-N^{k}_{A_2})}.
\end{equation}
To obtain the value of \([R_n]_Q\) one must repeat this procedure and average over a quantity which is equal to \(r_n\) if $q$ equals the required imbalance sector $q=Q$ and is \(0\) otherwise. 
In other words,
\begin{equation}\label{protequal}
[R_n]_Q=\Tr \{ \rho_A^{\otimes n} \hat{f}^{\mathrm{neg}}_{Q}(\{ \tilde{N} \} ) \},
\end{equation}
where the right hand side reflects the measurement protocol in the occupation basis $\tilde{N}$ after the Fourier transformation, and
\begin{equation}
f^{\mathrm{neg}}_{Q}(\{N\})\equiv\delta_{Q,\frac{1}{n}\sum_{k}(N^{k}_{A_1}-N^{k}_{A_2})}  \times e^{\frac{2\pi i}{n} \sum_{k}k(N^{k}_{A_1}-N^{k}_{A_2})}.
\end{equation}

Few remarks on our protocol are in order. Firstly, it is not necessary to perform the Fourier transform exclusively on the sites of subsystem $A$. Instead, one needs only to perform a Fourier transform on any region \(C\) of the system that contains region $A$. This simplification allows for experimental flexibility.
Secondly, if one measures the total particle numbers \(N^{k}_{A_1},N^{k}_{A_2}\) in subsystems \(A_1,A_2\) by performing occupation measurements on every site, one can immediately get the \Renyi{} negativity for all partitions of \(A\), \textit{i.e.} for all \(A'_1\cup A'_2 = A \subseteq C\). Thirdly, by evaluating \(Q\), the occupancy measurements automatically decompose the entanglement negativity into imbalance sectors.

We also note that we have restricted out analysis to bosons. The case of fermions was addressed for the second  \Renyi{}  entropy in Ref.~[\onlinecite{pichler2013thermal}] and we leave generalizations to future work~\cite{CESEGM}. In addition, while we only discussed the $n \ge 2$  \Renyi{}  entropies and negativities, one may use the methods in Ref.~[\onlinecite{gray2017measuring}] to access the corresponding entanglement entropy or negativity obtained by analytic continuation and taking the limit of $n \to 1$.

The proof of Eq.~(\ref{protequal}) follows from the relation
\({\Tr \{  [\rho_A]_Q^{\otimes n}\hat{T} \} = \Tr \{ \rho_A^{\otimes n} \hat{f}^{\mathrm{neg}}_{Q}(\{ \tilde{N} \} ) \}}\); the left hand side is the expectation value of the twisted swap operator in the $n$-copy state of the imbalance-$Q$ sector ${[\rho_A]_Q \equiv ([\rho_{A}^{\Tt}]_Q )^{\Tt}}$, which satisfies \({[R_n]_Q = \Tr \{  [\rho_A]_Q^{\otimes n}\hat{T} \}}\).
In order to show this relation we write a general state using the super-index notation,
\begin{equation}
|\mathbf{N} \rangle \equiv |N^{1}_{A_1},\dots,N^{n}_{A_1} \rangle^{i^{1}_{A_1}, \dots, i^{n}_{A_1}} |N^{1}_{A_2},\dots, N^{n}_{A_2} \rangle^{i^{1}_{A_2}, \dots, i^{n}_{A_2}},
\end{equation}  
where $\{ | N^{k}_{A_\alpha} \rangle^1,| N^{k}_{A_\alpha} \rangle^2,| N^{k}_{A_\alpha} \rangle^3,\dots \}$ span the Hilbert space of subsystem $A_\alpha$ in copy $k$. We proceed exactly along the lines of Eqs.~(\ref{step1}),~(\ref{step2}), and ~(\ref{step3}), and hence only provide key remarks. We use the fact that the total number of particles in each region $A_1$ and $A_2$ remains invariant under the Fourier transformation, ${\sum_{k=1}^n \tilde{N}^{k}_{A_\alpha} = \sum_{k=1}^n N^{k}_{A_\alpha}}$. This allows one to extract the delta function operator from the matrix element ${\langle \mathbf{N}' | \delta_{Q,\frac{1}{n}\sum_{k=1}^n(\tilde{N}^{k}_{A_1}-\tilde{N}^{k}_{A_2})}  \times e^{\frac{2\pi i}{n}\sum_{k=1}^{n}k(\tilde{N}^{k}_{A_1}-\tilde{N}^{k}_{A_2})} | \mathbf{N} \rangle}$.  Finally, we use the relation Eq.~(\ref{EqT})  to obtain a matrix element of the twisted swap operator $\langle \mathbf{N}' |\hat{T} | \mathbf{N} \rangle$. The twisted swap operator acts as
\begin{equation}
\hat{T}\ket{\begin{aligned}&N^{1}_{A_1},\dots,N^{n}_{A_1}\\ &N^{1}_{A_2},\dots,N^{n}_{A_2}\end{aligned}}^{\!\!\!\{i\}}=\ket{\begin{aligned}&N^{n}_{A_1}, N^{1}_{A_1},\dots,N^{n-1}_{A_1}\\ &N^{2}_{A_2},\dots,N^{n}_{A_2},N^{1\phantom{-1}}_{A_2}\end{aligned}}^{\!\!\!\{i\}}.
\end{equation}
States contributing to this matrix element satisfy ${N'^{k}_{A_1}=N^{k-1}_{A_1}}$ and  ${N'^{k}_{A_2}=N^{k+1}_{A_2}}$. In addition, charge conservation implies ${N^{k}_{A_1}+N^{k}_{A_2}=N'^{k}_{A_1}+N'^{k}_{A_2}}$.  This set of equations allows us to define
\begin{multline}\label{QNNNN}
Q\equiv N^{1}_{A_1}-N^{2}_{A_2}=N^{n}_{A_1}-N^{1}_{A_2}\\=N^{n-1}_{A_1}-N^{n}_{A_2}=\dots=N^{2}_{A_1}-N^{3}_{A_2}.
\end{multline}
By summing all these equations, one gets
\begin{equation}
Q=\frac{1}{n}\sum_{k=1}^n (N^{k}_{A_1} -N^{k}_{A_2})=\frac{1}{n}\sum_{k=1}^n (\tilde{N}^{k}_{A_1} -\tilde{N}^{k}_{A_2}).
\end{equation}
By observing Eq.~(\ref{QNNNN}) we can see that the states which contribute are exactly the blocks of the original density matrix that satisfy Eq.~(\ref{blocks1}), these precisely form the imbalance-$Q$ sector.

\section{Field Theory Analysis}\label{sec:cft}
Having shown the possibility to experimentally measure the separation of negativity into symmetry sectors, we now study one dimensional (1D) critical systems where general results for this quantity can be readily obtained. In such 1D critical systems the entanglement entropy shows the famous logarithmic scaling with the subsystem size, $S = \frac{c}{3} \log (\ell_A)+const.$, which can be decomposed into charge  sectors~\cite{goldstein2017symmetry}, $S = \sum_{N_A} [S]_{N_A}$; The contributions  $[S]_{N_A}$ were found~\cite{goldstein2017symmetry} to share the same distributions as the charge $N_A$ in region $A$~\cite{song2012bipartite}. We will now address a similar question for the negativity and its imbalance-decomposition.

Similar to the entanglement entropy scaling result, the negativity of two subsystems consisting of two adjacent intervals \(A_1,A_2\) , of lengths $\ell_1,\ell_2$, out of an infinite system in the ground state, has also been studied in the scaling limit. It acquires a universal form~\cite{calabrese2012negativity}, $\Tr\big|\rho^{\Tt}_A\big|=R_{n_e\to 1}  \propto  \big(\tfrac{\ell_1 \ell_2}{\ell_1 + \ell_2} \big)^{c/4}$, depending only on the central charge, $c$. We shall decompose this result into imbalance sectors. We note that although the case of of non-adjacent intervals may be treated using similar methods, it is more technically involved, and so we do not explicitly address it in this section.

\begin{figure}[t]
\includegraphics[width=1\linewidth]{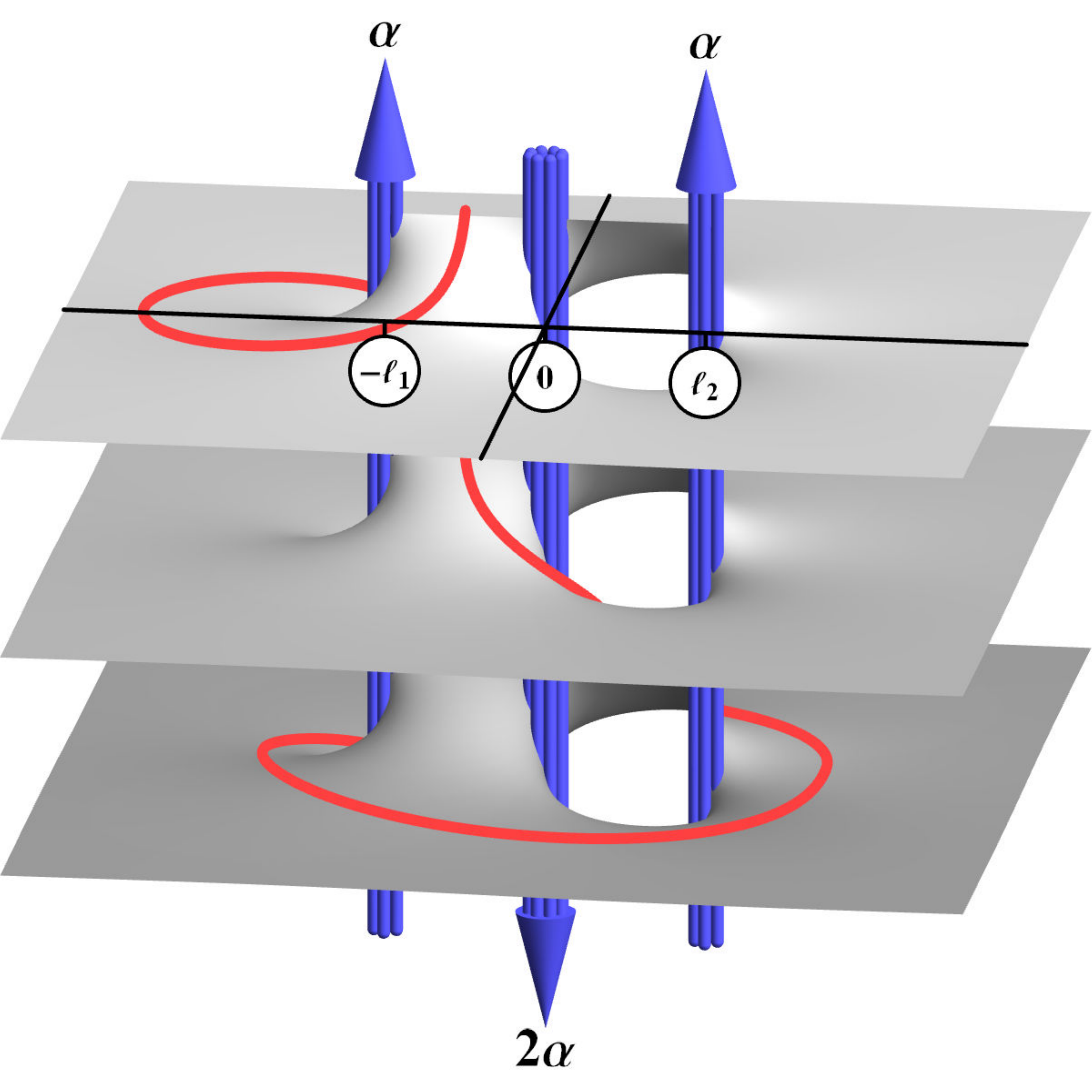}
\caption{(color online) Schematic representation of the \Renyi{} negativity \(\Tr\{(\rho_A^{\Tt})^n\}\) as an \(n\)-sheet Riemann surface with \({A_1=[-\ell_1,0]}\) and \({A_2=[0,\ell_2]}\). The phase factor, \(\exp\{i\alpha(N_1-N_2^{\Tt})\}\), is implemented by flux insertions \(\mathcal{V}_{\alpha}(-\ell_1)\mathcal{V}_{-2\alpha}(0)\mathcal{V}_{\alpha}(\ell_2)\) and represented by the vertical arrows. The winding line serves only as a visual aid.\label{fig:flux}} 	 	
\end{figure} 

We begin by briefly recapitulating the computation method of negativity based on Ref.~[\onlinecite{calabrese2012negativity}] using CFT. We note that these field theory methods are closely related in spirit to the $n$-copy construction of the system discussed in the previous section. In field theory\cite{calabrese2012negativity}, the \Renyi{} negativity is treated as a partition function of an \(n\)-sheet Riemann surface depicted in Fig.~\ref{fig:flux}. It is found to be determined by the 3-point correlation function of local ``twist" fields, 
\begin{equation}
R_n=\Tr\{(\rho_A^{\Tt})^n\} = \langle \mathcal{T}_n(-\ell_1) \tilde{\mathcal{T}}_n^2(0) \mathcal{T}_n(\ell_2) \rangle.
\end{equation}
The twist fields, \(\mathcal{T}_n\), generate the \(n\)-sheet Riemann surface depicted in Fig.~\ref{fig:flux}, and have scaling dimensions,
\begin{equation}\label{deltatau}
\Delta_{\mathcal{T}_n}=\frac{c(n-1/n)}{24},\quad
\Delta_{\mathcal{T}_{n_o}^2}=\Delta_{\mathcal{T}_{n_o}},\quad
\Delta_{\mathcal{T}_{n_e}^2}=2\Delta_{\mathcal{T}_{n_e/2}}.
\end{equation}
Here, one splits the results for the scaling dimension of the squared twist field for even (\(n=n_e\)) and odd (\(n=n_o\)) cases. Using these scaling dimensions, the desired 3-point function is easily evaluated,
\begin{gather}
R_{n_o} \propto (\ell_1 \ell_2(\ell_1 + \ell_2))^{-\frac{c}{12}\left(\scriptscriptstyle{n_o}-\frac{1}{n_o}\right)},\\
R_{n_e} \propto (\ell_1 \ell_2)^{-\frac{c}{6}\left(\frac{n_e}{2}-\frac{2}{n_e}\right)}(\ell_1 + \ell_2)^{-\frac{c}{6}\left(\frac{n_e}{2}+\frac{1}{n_e}\right)},\\
\Tr\big|\rho^{\Tt}_A\big|=R_{n_e\to 1}  \propto  \left(\frac{\ell_1 \ell_2}{\ell_1 + \ell_2} \right)^{\frac{c}{4}}.
\end{gather}

We herein implement the negativity splitting of Eq.~(\ref{resolve1}). To do so, we use the Fourier representation of the projection operator,
\begin{equation}\label{projP}
\hat{P}_Q
=\int_{-\pi}^\pi\frac{d\alpha}{2\pi}e^{-i\alpha Q}e^{i\alpha(\hat{N}_1^{\phantom{|}}-\hat{N}_2^{\Tt})},
\quad \sum_Q \hat{P}_Q = 1.
\end{equation}
In the context of field theory, however, one treats the system in the continuum limit such that
\begin{equation}\label{ProjCFT}
\hat{P}_Q
=\int_{-\infty}^\infty\frac{d\alpha}{2\pi}e^{-i\alpha Q}e^{i\alpha(\hat{N}_1^{\phantom{|}}-\hat{N}_2^{\Tt})},
\quad \int_{-\infty}^{\infty} dQ \hat{P}_Q  = 1.
\end{equation}

We now restrict our attention to a CFT of central charge $c=1$  which is equivalent to 1D massless bosons and thus to Luttinger liquids (gapless interacting 1D fermions~\cite{giamarchi2003quantum,gogolin2004bosonization}). Applying the methods of Ref.~[\onlinecite{goldstein2017symmetry}], the phase factor \(e^{i \alpha (\hat{N}_1-\langle \hat{N}_1\rangle)}\) may be implemented by two vertex operators \(\mathcal{V}_\alpha\) at \({z=-\ell_1}\) and  \(\mathcal{V}_{-\alpha}\) at \({z=0}\). Moreover, within the CFT's geometrical basis, \(\hat{N}\) is a real operator and thus \({\hat{N}^{\Tt}=\hat{N}^\ast=\hat{N}}\) and so a similar vertex operator insertion may account for \(e^{-i \alpha (\hat{N}_2-\langle \hat{N}_2\rangle)}\). In general, such insertions may be done in any of the \({k=1\dots n}\) sheets with different phases $\alpha_k$, However, these vertex operators may be interpreted as a flux insertion, in which case, gauge invariance implies that only the overall flux has physical implications. Indeed, using the techniques of Refs.~[\onlinecite{Belin2013},\onlinecite{Cornfeld2017Entanglement}], we show in Appendix~\ref{app:cft}, that this physical intuition holds, and 
that the correlations depend only on the total flux \({\alpha=\sum_{k=1}^n \alpha_k}\). Considering a generic Luttinger
liquid with parameter $K$, we find the scaling dimension of the fluxed twist operator \((\mathcal{T}_n\mathcal{V}_\alpha)\) to be~\cite{goldstein2017symmetry}
\begin{equation}\label{deltaalpha}
\Delta_n(\alpha) = \frac{1}{24}\left(n-\frac{1}{n}\right)+\frac{K}{2n}\left(\frac{\alpha}{2 \pi}\right)^2.
\end{equation}
In terms of these fluxed twist operators, the negativities are also related to 3-point functions,
\begin{align}
[R_n]_Q&=\int_{-\infty}^\infty\frac{d\alpha}{2\pi}e^{-i\alpha (Q-\langle \hat{Q}\rangle)}R_n(\alpha)\label{Rna}\\
R_n(\alpha)&=e^{-i\alpha\langle\hat{N}_1^{\phantom{|}}-\hat{N}_2^{\Tt}\rangle}\Tr\{e^{i\alpha(\hat{N}_1^{\phantom{|}}-\hat{N}_2^{\Tt})}(\rho_A^{\Tt})^n\} \nonumber\\
&= \langle (\mathcal{T}_n \mathcal{V}_\alpha)_{z=-\ell_1}(\tilde{\mathcal{T}}_n^2\mathcal{V}_{-2\alpha})_{z=0} (\mathcal{T}_n\mathcal{V}_\alpha)_{z=\ell_2} \rangle.
\end{align}
Using the scaling dimensions given in Eqs.~(\ref{deltatau}),~(\ref{deltaalpha}), one evaluates
\begin{equation}
\frac{R_n(\alpha)}{R_n}\propto(\ell_1\ell_2)^{-\frac{4K}{n}\left(\frac{\alpha}{2 \pi}\right)^2}(\ell_1+\ell_2)^{\frac{2K}{n}\left(\frac{\alpha}{2 \pi}\right)^2}.
\end{equation}
Upon integration over $\alpha$ in Eq.~(\ref{Rna}) we obtain the result,
\begin{gather}
\frac{[R_n]_Q}{R_n}=\sqrt{\frac{\pi n}{2K\log\frac{\ell_1^2\ell_2^2}{(\ell_1+\ell_2)\Lambda^{3}}}}\exp\Bigg(-\frac{n \pi^2 (Q-\langle\hat{Q}\rangle)^2}{2K\log\frac{\ell_1^2\ell_2^2}{(\ell_1+\ell_2)\Lambda^{3}}}\Bigg),\label{CFTres1}
\\
\frac{\Tr\big|[\rho_A^{\Tt}]_Q\big|}{\Tr\big|\rho_A^{\Tt}\big|}=\left[\frac{[R_n]_Q}{R_n}\right]_{n_e\to1},\label{CFTres2}
\end{gather}
where \({\Lambda\sim a \ll\ell}\) is a short distance cut-off, and \(a\) is the lattice spacing. These equations provide an analytic expression for the imbalance-resolved \Renyi{} negativities and negativity which are numerically inaccessible for large systems. 

One may simply cast the result for the negativity Eqs.~(\ref{CFTres1}),~(\ref{CFTres2}), as
\begin{equation}\label{CFTres3}
\Tr\big|[\rho_A^{\Tt}]_Q\big|=\langle\hat{P}_{N_1-N_2=Q}\rangle\Tr\big|\rho_A^{\Tt}\big|.
\end{equation}
It follows from the identity \({\Tr \hat{O}=\Tr\{\hat{O}^\Tt\}}\) which leads to \({[R_1]_Q=\langle\hat{P}_{N_1-N_2=Q}\rangle}\); see Eq.~(\ref{projO}). This result signifies that the imbalance-resolved negativity depends only on the probability distribution function of the occupation imbalance $N_1 - N_2$ itself, $\langle \hat{P}_{N_1 - N_2=Q} \rangle $.  The latter depends on the Luttinger parameter as seen in Eq.~(\ref{CFTres1}).

The expression for the \(n=3\) \Renyi{} negativity, is numerically accessible and is used in the following subsection to validate our results; see Fig.~\ref{fig:fit}.

\subsection{Numerics}\label{sec:num}

\begin{figure}[t]
\includegraphics[width=1\linewidth]{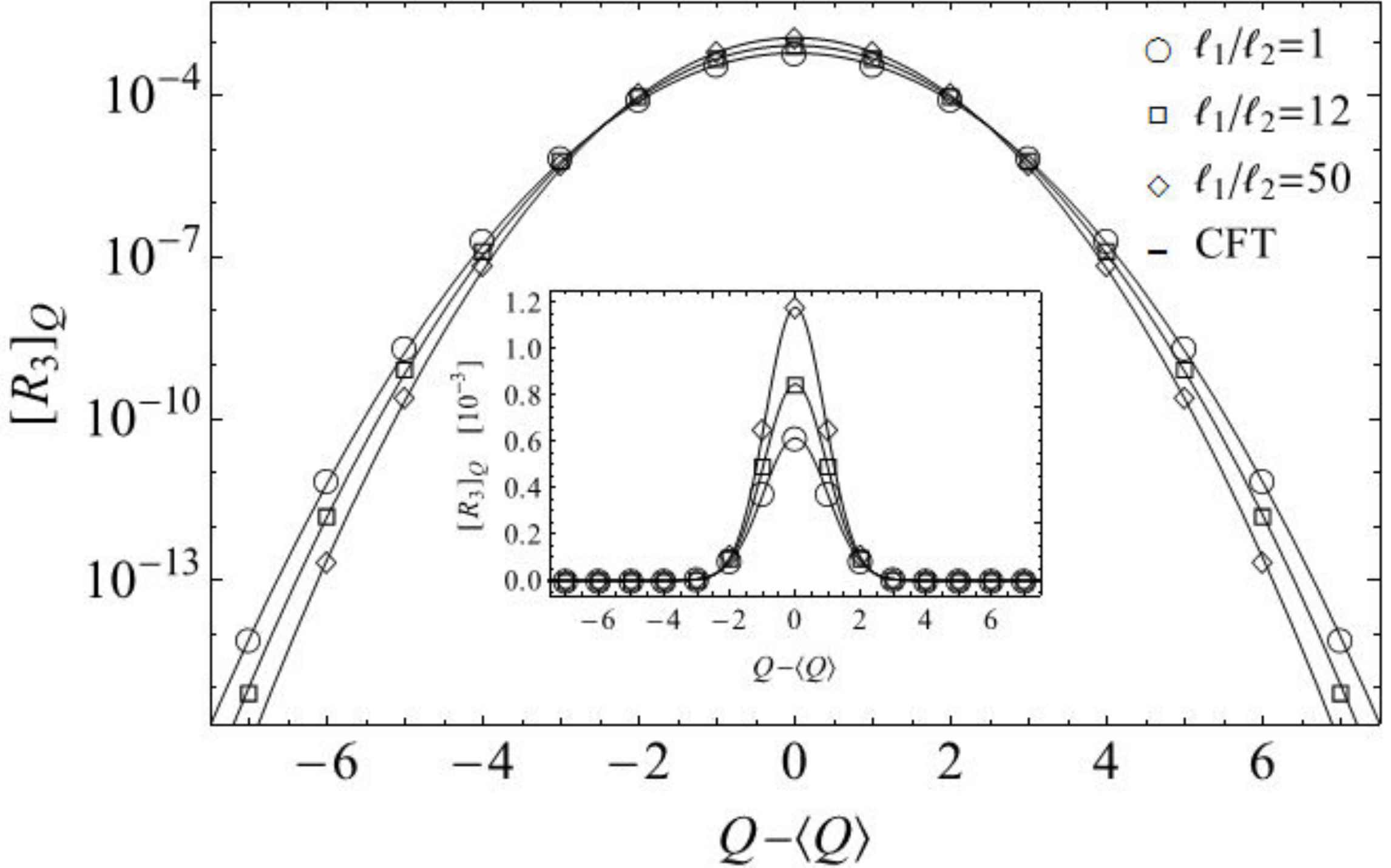}
\caption{Exact numerical calculations of \([R_3]_Q\) for a half-filled tight-binding chain of free fermions are displayed on a logarithmic scale, and fitted to the CFT predictions (continuous lines) of Eq.~(\ref{CFTres1}) with \(K=1\). The inset shows the same data displayed on a linear scale. All graphs are for \(\ell_1+\ell_2=4000a\), and infinite-length environment \(B\). \label{fig:fit}} 	 	
\end{figure} 

For adjacent intervals \(A_1,A_2\), one may use the Jordan-Wigner transformation to relate the entanglement of either a bosonic spin-half chain or hard-core bosons, to that of free fermions. This corresponds to the case of Luttinger parameter ${K=1}$, whereby we may use free fermions methods to effectively calculate the negativities.

By generalizing the analysis of Refs.~[\onlinecite{eisler2015partial},\onlinecite{Eisler2016Entanglement}] and using the results of Refs.~[\onlinecite{Klich2002FCS},\onlinecite{Klich2014FCS}], one may study the \Renyi{} negativity in any equilibrium free fermion system with density matrix \({\rho\propto e^{-\beta\hat{H}}}\) via a double-gaussian representation. Specifically, one may calculate \(\Tr\{ e^{\hat{O}}(\rho_A^{\Tt})^n\}\) for any operator \(\hat{O}\) that is quadratic in fermionic creation and annihilation operators; see Appendix~\ref{app:numerics}. To utilize these techniques, we use Eq.~(\ref{projP}) with  \({\hat{O}=i\alpha(\hat{N}_1^{\phantom{|}}-\hat{N}_2^{\Tt})}\),
\begin{equation}\label{projO}
\begin{aligned}
{[R_n]_Q}&=\Tr\big\{\hat{P}_Q(\rho_A^{\Tt})^n\big\}\\
&=\int_{-\pi}^\pi\frac{d\alpha}{2\pi}e^{-i\alpha Q}\Tr\big\{e^{i\alpha(\hat{N}_1^{\phantom{|}}-\hat{N}_2^{\Tt})}(\rho_A^{\Tt})^n\big\}.
\end{aligned}
\end{equation}

Numerical results for ${[R_3]_Q}$ for free fermions at zero temperature are shown in Fig.~\ref{fig:fit}, and are fitted to the CFT predictions of Eq.~(\ref{CFTres1}) with $K=1$. Analyses for \(\Lambda/a\) of similar systems are known~\cite{Jin2004,song2012bipartite} to range between \(0.1017\) and \(0.1033\). Within the validity regime \({\Lambda\ll\ell}\) the functional dependence on \(\Lambda\) is negligible, and all values in the aforementioned range yield excellent fits to the data; the figure is plotted with \({\Lambda/a=0.}1\). 
Further details about the numerical technique are given in Appendix~\ref{app:numerics}.

\section{Conclusions and outlook}  
\label{sec:conclusion}
We studied entanglement negativity  in general many-body systems possessing a global conserved charge, and found it to be decomposable into symmetry sectors. Interestingly, due to the partial transposition operation involved in the definition of negativity, the resulting operator that commutes with the partially transposed density matrix is not the total charge, but rather an imbalance operator which is essentially the particle number difference between two regions.

We have proposed an experimental protocol for the measurement of these contributions to the \Renyi{} negativities using existing cold-atom technologies. While current cold-atom detection schemes~\cite{Islam2015Measuring} are based on measurements of the \emph{parity} of the on-site occupation due to unavoidable two-atom molecule formation, the measurements of $n>2$ \Renyi{} entropies proposed here require full integer occupation detection. This requirement may be relaxed for hard-core interacting bosons, or specificity for \emph{fermions}. An issue
which we have not addressed here is the entanglement
and negativity measurement protocols for fermions where additional fermionic exchange phases should be taken into account~\cite{CESEGM}.  

We have also attained field theory predictions for the distribution of entanglement in critical 1D systems, and have verified them numerically. In addition to critical systems, one may study the symmetry decomposition of negativity in gapped systems. It would be interesting to further explore physical consequences of this imbalance decomposition of negativity in topological systems~\cite{PhysRevA.88.042318,PhysRevA.88.042319,wen2016topological}.

\emph{Acknowledgements}  	  
E.S. was supported in part by the Israel Science Foundation (Grant No. 1243/13) and by the US-Israel Binational Science Foundation (Grant No. 2016255). M.G. was supported by the Israel Science Foundation (Grant No. 227/15), the German Israeli Foundation (Grant No. I-1259-303.10), the US-Israel Binational Science Foundation (Grant No. 2016224), and the Israel Ministry of Science and Technology (Contract No. 3-12419).

\appendix
\section{Fluxed Twist Operators}\label{app:cft}
\renewcommand{\theequation}{A\arabic{equation}}
\setcounter{equation}{0}
In this appendix we investigate the twisted flux operator \((\mathcal{T}_n\mathcal{V}_{\alpha})\) of Sec.~\ref{sec:cft}. Following Refs.~[\onlinecite{goldstein2017symmetry},\onlinecite{Belin2013},\onlinecite{Cornfeld2017Entanglement}], we find its scaling dimension Eq.~(\ref{deltaalpha}),  and show that it depends only on the total flux insertion.
For simplicity, we set \({K=1}\) (free fermions) within this appendix.

A vertex operator insertion of \(\mathcal{V}_{\alpha_k}\) at copy \(k=1\dots n\) creates a monodromy of \(\alpha\) when crossing to the next copy. Therefore, the boson fields \(\phi_k\) satisfy the following relations upon crossing the \(A_1\) cut at \([-\ell_1,0]\)
\begin{equation}
\Psi\mapsto T_{\{\alpha\}}\Psi,
\end{equation}
where the field vector \(\Psi\) and transformation matrix \(T_{\{\alpha\}}\) satisfy~\cite{Casini2005Entanglement}.
\begin{equation}
\Psi=
\left(\begin{smallmatrix}
e^{i\phi_1}\\
e^{i\phi_2}\\
\vdots\\
e^{i\phi_n}
\end{smallmatrix}\right),\quad
T_{\{\alpha\}}=
\left(\begin{smallmatrix}
0 						&e^{i\alpha_1}	& 				& 		\\
&0				&e^{i\alpha_2}	& 		\\
&				&\ddots			&\ddots \\
(-1)^{n+1} e^{i\alpha_n}	&				&				&0
\end{smallmatrix}\right).
\end{equation}
This transformation matrix has eigenvalues
\begin{equation}
\lambda_p=e^{i\frac{1}{n}\sum_{k=1}^n\alpha_k}e^{2\pi i \frac{p}{n}},\qquad p=-\tfrac{n-1}{2}\dots\tfrac{n-1}{2}.
\end{equation}
On the other hand upon crossing the \(A_2\) cut at \([0,\ell_2]\) the relation reverses
\begin{equation}
\Psi\mapsto T_{\{-\alpha\}}^{\mathrm{T}}\Psi.
\end{equation}
Since \({T_{\{-\alpha\}}^{\mathrm{T}}=T_{\{\alpha\}}^\dagger=T_{\{\alpha\}}^{-1}}\), one has \({[T_{\{-\alpha\}}^{\mathrm{T}},T_{\{\alpha\}}^{\phantom{|}}]=0}\). This enables one to simultaneously diagonalize the transformations \(T_{\{\alpha\}}^{\phantom{|}},T_{\{-\alpha\}}^{\mathrm{T}}\) in the cuts \(A_1,A_2\). This implies that the \(p\)~basis eigenvector fields \(\phi_p\) remain decoupled and so all correlations depend only on \({\alpha=\sum_{k=1}^{n}\alpha_k}\). In agreement with the monodromies, one may decompose~\cite{Belin2013,Cornfeld2017Entanglement} the fluxed twist operator \({(\mathcal{T}_n\mathcal{V}_{\alpha})=\prod_p e^{i(\frac{p}{n}+\frac{\alpha}{2\pi n})\phi_p}}\) and find its scaling dimension
\begin{equation}
\Delta_n(\alpha) = \tfrac{1}{2}\sum_{p}\left(\tfrac{p}{n}+\tfrac{\alpha}{2\pi n}\right)^2=\tfrac{1}{24}\left(\scriptstyle{n}-\tfrac{1}{n}\right)+\tfrac{1}{2n}\left(\tfrac{\alpha}{2 \pi}\right)^2.
\end{equation}
Similar scaling dimensions may be found for the other fluxed twist operators.

\section{Numerical Technique}\label{app:numerics}
\renewcommand{\theequation}{B\arabic{equation}}
\setcounter{equation}{0}
In this appendix we briefly review the double-gaussion representation of Refs.~[\onlinecite{eisler2015partial},\onlinecite{Eisler2016Entanglement}] and use it to explicitly present our numerical procedure of Eq.~(\ref{projO}) which are displayed in Fig.~\ref{fig:fit}.

Refs.~[\onlinecite{eisler2015partial},\onlinecite{Eisler2016Entanglement}] have shown that the partially transposed density matrix of free electron systems may be written as a sum of Gaussian matrices,
\begin{gather}
\rho_A^{\Tt}=\sum_{\sigma=\pm}u_\sigma\frac{\hat{O}_\sigma}{\Tr\hat{O}_\sigma},\\
\hat{O}_\sigma=e^{\sum_{ij}\hat{c}^\dagger_i [W_\sigma]_{ij} \hat{c}_j},\qquad u_\sigma=\tfrac{1}{\sqrt{2}}e^{-i\frac{\pi}{4}\sigma}.
\end{gather}
The \(W_\sigma\) matrices are related to the fermionic green function \({{[C]_{ij}=\langle[\hat{c}^\dagger_i,\hat{c}_j]\rangle}}\) by
\begin{equation}\label{Gdef}
e^{W_\sigma}=\frac{1+G_\sigma}{1-G_\sigma},\qquad G_\sigma=\left(\begin{array}{c|c}
C^{11} &  \sigma i C^{12}\\
\hline
\sigma i C^{21} & C^{22}\\
\end{array}\right),
\end{equation} 
where \(C^{\alpha\beta}\) are the blocks of \(C\) in regions \(A_\alpha,A_\beta\).

To obtain Eq.~(\ref{projO}), we first study a generic quadratic operator \({\hat{X}=\sum_{ij}\hat{c}^\dagger_i [X]_{ij} \hat{c}_j}\),
\begin{equation}
\Tr\{ e^{\hat{X}}(\rho_A^{\Tt})^n\}=\sum_{\{\sigma\}}u_{\{\sigma\}}\frac{\Tr\{e^{\hat{X}}\prod_i\hat{O}_{\sigma_i}\}}{\prod_i\Tr \hat{O}_{\sigma_i}},
\end{equation}
where \({u_{\{\sigma\}}=2^{-\frac{n}{2}}e^{-i\frac{\pi}{4}\sum_i\sigma_i}}\) are the coefficients of \(\{\hat{O}_\sigma\}\) in the expansion of \((\rho_A^{\Tt})^n\). We use the results of Refs.~[\onlinecite{Klich2002FCS},\onlinecite{Klich2014FCS}] to evaluate
\begin{equation}
\Tr\{ e^{\hat{X}}(\rho_A^{\Tt})^n\}=\sum_{\{\sigma\}}u_{\{\sigma\}}\frac{\det(1+e^X\prod_i e^{W_{\sigma_i}})}{\prod_i\det(1+e^{W_{\sigma_i}})}.
\end{equation}
When \({[X,W_\sigma]=0}\), one may use Eq.~(\ref{Gdef}) and some matrix algebra to get
\begin{multline}\label{3p1}
\Tr\{ e^{\hat{X}}(\rho_A^{\Tt})^3\}=-\frac{1}{2}\det\left(\left(\tfrac{1-G_+}{2}\right)^3+e^X\left(\tfrac{1+G_+}{2}\right)^3\right)\\
+\frac{3}{2}\det\left(\left(\tfrac{1-G_+}{2}\right)^2\tfrac{1-G_-}{2}+e^X\left(\tfrac{1+G_+}{2}\right)^2\tfrac{1+G_-}{2}\right).
\end{multline}
This may straightforwardly be numerically estimated.

As noted in Sec.~\ref{sec:num}, for the calculation of \(R_3\) we pick \({\hat{O}=i\alpha(\hat{N}_1-\hat{N}_2^{\Tt})}\) in Eq.~(\ref{projO}). However, in this basis~\cite{eisler2015partial} one has \({\hat{N}_2^{\Tt}=L_2-\hat{N}_2}\) where \(L_2\) is the number of sites in \(A_2\). Therefore,  \({\hat{O}=i\alpha(\hat{N}-L_2)}\) and
\begin{equation}
[R_n]_Q=\int_{-\pi}^\pi\frac{d\alpha}{2\pi}e^{-i\alpha Q-i\alpha L_2}\Tr\big\{e^{i\alpha\hat{N}}(\rho_A^{\Tt})^n\big\},
\end{equation}
such that \({\hat{N}=\sum_{ij}\hat{c}^\dagger_i[\mathbb{I}]_{ij}\hat{c}_j}\) clearly satisfies \({[\mathbb{I},W_\sigma]=0}\), and one may use Eq.~(\ref{3p1}).

\end{document}